\documentclass[manuscript]{aastex61}

\usepackage{CJK}
\usepackage{amssymb}
\usepackage{multirow}
\usepackage{morefloats}
\usepackage{amsmath}
\usepackage{bigstrut}
\usepackage{booktabs}
\usepackage{natbib}
\usepackage{float}
\usepackage{graphicx,epsfig,fancyhdr,epsf,txfonts,epstopdf}
\usepackage{latexsym,bbm}
\usepackage{lineno}
\usepackage{color}
\usepackage{ulem}
\usepackage{threeparttable}
\usepackage{longtable}

\received{***}
\revised{***}
\accepted{by APJ - 04-Jan-2021}

\shortauthors{Yousefzadeh et al.}
\shorttitle{ECME Inside A Coronal Loop}

\begin{document}

%\begin{CJK*}{UTF8}{gbsn}
\title{Harmonic ECME Excited by Energetic Electrons Travelling Inside A Coronal Loop}

\author{Mehdi Yousefzadeh}\thanks{E-mail: m.yousefzadeh6@gmail.com}
\affil{Institute of Space Sciences, Institute of Frontier and Interdisciplinary Science, Shandong University, Shandong, China.\\}

\author{Hao Ning}
\affiliation{Institute of Space Sciences, Institute of Frontier and Interdisciplinary Science, Shandong University, Shandong, China.\\}

\author{Yao Chen}%\thanks{E-mail: yaochen@sdu.edu.cn}
\affiliation{Institute of Space Sciences, Institute of Frontier and Interdisciplinary Science, Shandong University, Shandong, China.\\}

\correspondingauthor{Yao Chen}
\email{yaochen@sdu.edu.cn}

\begin{abstract}
A complete understanding of solar radio bursts requires developing numerical techniques which can connect large-scale activities with kinetic plasma processes. As a starting point, this study presents a numerical scheme combining three different techniques: (1) extrapolation of magnetic field overlying a specific active region in order to derive the background field, (2) guiding-center simulation of dynamics of millions of particles within a selected loop to reveal the integral velocity distribution function (VDF) around certain sections of the loop, and (3) particle-in-cell (PIC) simulation of kinetic instabilities driven by energetic electrons initiated by the obtained distributions. Scattering effects at various levels (weak, moderate, and strong) due to wave$\backslash$turbulence-particle interaction are considered using prescribed time scales of scattering. It was found that the obtained VDFs contain strip-like and loss-cone features with positive gradient, and both features are capable of driving
electron cyclotron maser emission (ECME), which is a viable radiation mechanism for some solar radio bursts, in particular, solar radio spikes.
The strip-like feature is important in driving the harmonic X mode, while the loss-cone feature can be important in driving the fundamental X mode. In the weak-scattering case, the rate of energy conversion from energetic electrons to X2 can reach up to $\sim2.9 \times 10^{-3}$$E_{k_{0}}$, where $E_{k_{0}}$ is the initial kinetic energy of energetic electrons. The study demonstrates a novel way of exciting X2 mode in the corona during solar flares, and provides new sight into how escaping radiation can be generated within a coronal loop during solar flares.
\end{abstract}


\section{Introduction}
%\linenumbers
Solar radio bursts represent sudden enhancement of emission at radio wavelengths released from the solar atmosphere. Various types of bursts have been classified according to their manifestation on the dynamic spectrum, such as bursts of type I to V, millisecond solar spikes, flare decimetric continuum, etc (\citealp{1963_Wild};  \citealp{1985_Mclean}; see, e.g.,
 \citealp{2012_Feng}, \citealp{2017_Li}, \citealp{2014_Chen}, and \citealp{2019_Vasanth} for latest studies).
Among them, solar spikes are radio bursts closely associated with the impulsive stage of solar flares. Spikes are characterized by their extremely high brightness temperature of up to 10$^{15}$ K, the highest among all types of solar radio bursts. They appear in metric-decimetric wavelength, exhibiting very narrow-band spectral width with rapidly-rising intensity profile and extremely short characteristic time scales ($\sim$ a few to tens of ms for individual burst). Thousands of spikes can exist in an event. This leads some authors to suggest that solar spikes represent elementary energy release events during solar flares (see, e.g.,   \citealp{1986_Benz}; \citealp{1986a_Benz}; \citealp{1998_Fleishman}).

In earlier studies, ECME is considered to be the emission mechanism of solar spikes (see, e.g., \citealp{1980_Holman}; \citealp{1982_Melrose}; \citealp{1991_Robinson}). The ECME mechanism was initially proposed by \citet{1958_Twiss}, \citet{1959_Gaponov} and \citet{1959_Schneider}. \citet{1979_Wu} realized the importance of relativistic effect in the resonance condition, and applied ECME to Decametric Radiation (DAM) of Jupiter. Later, the theory was used to explain solar radio bursts, such as solar spikes (\citealp{1982_Melrose}; \citealp{1984_Sharma}; \citealp{1990a_Aschwanden}) and type IIIs (\citealp{2005_Wu}).
According to \citet{1979_Wu} and follow-up studies, ECME can be excited by energetic electrons with a positive gradient along the perpendicular direction in the velocity space, i.e., $\partial f /\partial {v_\perp}>0$, in plasmas with $\omega_{pe}/\Omega_{ce} <1$. In addition, ECME favors the growth of X mode. In plasmas with $\omega_{pe}/\Omega_{ce}>1$ the growth rates of X and O modes decline rapidly while Z (or the electrostatic upper hybrid mode at large wave number) and W become the dominant modes (\citealp{1984_Sharma}; \citealp{2009_Lee}; \citealp{2013_Yi}; \citealp{2020_Ni}).

Most earlier studies suggest the solar spikes are excited by ECME of energetic electrons with the loss-cone distribution, i.e., via the loss-cone maser. Such distribution can form when electrons are trapped and bounced within a coronal loop. Nevertheless, unresolved issues exist that hinder us from a complete understanding of the exact emission mechanism and how one can use radio data to infer coronal properties. One major difficulty lies in the fact that any solar radio bursts are a consequence of multi-scale process. An event would start from large-scale magnetic eruptive activities in the astronomical to MHD scale, followed by particle acceleration via magnetic reconnection or shocks in the particle kinetic scale, and excitation and propagation of radiations. Present computational resources are far from enough to simulate such cross-scale physics.

Another difficulty is that the loss-cone maser for solar spikes mainly excite the fundamental X mode (X1), and it is difficult for X1 to pass through the second harmonic absorption layer during its escape from the source (\citealp{1982_Melrose}).
This gives the escaping difficulty of ECME when being applied to solar radio bursts. To resolve this, \citet{2016_Melrose} proposed that a low-density cavity exists in the source region, which can duct the X mode out. Existence of such low-density cavity in the corona has not been verified yet. Another option is to have direct excitation of X at higher harmonics such as X2 and X3, since the absorption rates at the third or fourth harmonic layer are considerably less in comparison to that at the second harmonic. Yet, direct and efficient excitation of X2 and X3 are rarely reported.

Most, if not all, existing studies on emission mechanism of solar radio bursts, including solar spikes, started from analytically-prescribed distribution function of energetic electrons. Such function is not determined by the particle dynamics in a self-consistent way. In this study, we present a preliminary effort to connect macro-scale dynamics of energetic electrons with micro-scale kinetic instabilities energized by these electrons to explain the possible emission mechanisms behind solar spikes. This is done by combining three sets of numerical techniques working at different scales. It was found that the VDFs deduced from the simulations contain certain interesting features that are capable of exciting emission efficiently at the second harmonic (X2) via ECME, thus providing a novel approach to resolve the escaping difficulty of ECME. The following section introduces numerical techniques used here, the third section presents analysis and major results, summary and discussion are given in the last section.

\section{Numerical Schemes}
To bridge the physics of different spatial scales and simplify the effort as much as possible, we combine three sets of numerical techniques, including (1) the nonlinear force-free field (NLFFF) extrapolation method (\citealp{2004_Wiegelmann} and \citealp{2006_Wiegelmann}) to describe large-scale topology of corona magnetic field within which electrons are traveling and trapped, (2) the guiding-center (GC) simplification method (e.g., \citealp{1963_Northrop}; \citealp{2010a_Gordovskyy}) to simulate dynamical motions of a large-number of electrons so as to infer evolution of their VDF, and (3) the particle-in-cell (PIC) simulation fed by the obtained VDFs to explore the excited kinetic instabilities and radiations. This represents a significantly-simplified approach for the proposed study.

\subsection{NLFFF Extrapolation of the Coronal Magnetic Field}
To start the simulation, we selected one specific active region (NOAA AR11283), which released a group of eruptions including three big flares (M5.3, X2.1, and X1.8) on September 6 and 7, 2011. It has been investigated by several authors (e.g., \citealp{2014_Ruan}; \citealp{2014_Liu}; \citealp{2015_Romano}).
Here we start from the magnetic condition before the X2.1 flare. The NLFFF extrapolation was based on the magnetogram data observed at 21:30 UT by the Helioseismic and Magnetic Imager (HMI; \citealp{2012_Schou}).

The HMI data and the result of extrapolation are shown in Figure 1(a), from which we observe a clear twisted flux rope structure whose eruption releases the X2.1 flare. Three representative field lines with different amount of twist are plotted. The loop shown in panel (b) contains no significant twist, the field strengths at the two foot-points are close to 1800 G and 1200 G, respectively. The location with the minimum field strength is close to the loop top. The loop in panel (c) presents the M-like topology with significant twist, the field strengths at the two foot-points are nearly 300 G and 1800 G, respectively. The location with the minimum field strength is located at the right foot-point. The field line in panel (d) has more twists, the field strength at its left foot is 1400 G, decreasing to a local minimum at 800 G and then increasing to the maximum at nearly 1300 G around the top part, then decreasing to 400 G and further increasing to 1200 G at the right foot.\\

As a starting point, we conduct simulation of particle dynamics within the simple untwisted loop configuration, i.e., that shown in panel (b). Note that selection of the specific AR and the loop is just to provide a realistic magnetic condition of the corona so as to initiate the simulation. The loop forms a simple magnetic trap (or mirror) and represents a typical field line in the background corona overlying the pre-eruption flux rope structure. Using other twisted configuration would be interesting yet more complex, and the essential physics should be revealed by the study using the simple untwisted loop structure. From the extrapolation technique we can obtain three-dimensional distribution of field strength and direction along the whole loop. The information will be fed to the following GC simulation of particle dynamics as input. The time scale of the bouncing motion can be estimated with the loop length and the average energy of electrons. Here, the height of the selected loop structure is about 100 arcsec, and the length is nearly 300 arcsec. Thus, it takes $\sim 3$s for electrons with an energy of 15 keV to travel such distance.

\subsection{Guiding-center simulation of dynamics of energetic electrons}
VDF of energetic electrons accounting for radio emission remains unknown. It is related to the acceleration and heating mechanism and further wave-particle interaction during solar flares. In any case, strong heating seems to be an intrinsic part of solar flares, therefore one may obtain heated electrons with high-temperature Maxwellian-like distribution. The heated electrons will be injected into a loop which might be the post-reconnection loop or one of the loops directly connected to the reconnection region. If reconnection takes place high in the corona, as indicated by X-ray data of flares (\citealp{1994_Masuda}; \citealp{2003_Sui}; \citealp{2017_Chen}), top of loops could be a probable place of injection. To simulate the following dynamic motion, we inject two million of electrons with a Maxwellian distribution around the loop top in an impulsive way (labelled with “I” in Figure 1(b)). The length of the injection section is $\sim$ 5 arcsec. The VDF is expressed by
\begin{equation}
f_0=\frac{1}{(2\pi)^{3/2}v_{0}^3}\exp\left(-\frac{u^2}{2v_0^2}\right)
\end{equation}
where $v_{0}=0.24$ c and c is the speed of light. The GC simplification is described with standard relativistic equations (\citealp{1963_Northrop}; \citealp{2010a_Gordovskyy}). To apply this method, the characteristic time and spatial scales of particle gyro-motion should be much less than those of the magnetic field variation. For the simple untwisted large-scale loop structure, such conditions are well satisfied. The classical Bulirsch-Stoer approach is employed to solve the GC equations, which is characterized by a high level of accuracy in energy conservation.

Pre-existing waves and those excited after injection can scatter electrons and modify their VDFs. To simulate such effect, we introduce a parameter $\tau_{s}$ to represent the time scale of pitch-angle scattering: within each interval of $\tau_{s}$, pitch angle of an electron will be changed to a random value in a range of [0, $\pi$]. A larger $\tau_s$ represents weaker scattering and vice versa. We refer to \citet{2007_Piana} and \citet{2013_Chen} for characteristic values of $\tau_{s}$ during solar flares. For electrons of few tens of Kev, $\tau_{s}$ could be around 0.1s to 1s. Thus, we set $\tau_s$ to be 5s, 1s, and 0.5s, to represent weak (Case W), moderate (Case M), and strong (Case S) levels of scattering, respectively.\\

\subsection{VPIC simulation of ECME driven by energetic electrons}
We use the Vector-PIC (VPIC) open-source code released by the Los Alamos National Labs for PIC simulation. The code employs a second-order, explicit, leapfrog algorithm to update particle positions and velocities, along with a second-order finite-difference time-domain solver for evolution of electric and magnetic fields which are described by the full Maxwell equations (\citealp{2008a_Bowers},\citealp{2008b_Bowers,2009_Bowers}).

To initiate the PIC simulation, we adopt Maxwellian distribution for background plasmas consisting of electrons and protons. The background magnetic field is set to be along $\vec{e}_z$ ($\vec B_0 = B_0 \vec{e}_z$) and the wave vector ($\vec k$) is within the $xOz$ plane. VDFs of energetic electrons are taken from the above GC simulation, and only those at the loop top are used. According to the NLFFF extrapolation the field strength around the loop top is $\sim$400 G, the number density of background plasmas $n_{0}$ is assumed to be $10^{9}$ cm$^{-3}$, thus the ratio of plasma oscillation frequency and electron gyro-frequency, i.e., $\omega_{pe}/\Omega_{ce}$, is 0.25. This means that ECME should be important. The density ratio of energetic and background electrons ($n_{e}/n_{0}$) is taken to be 0.05. The grid spacing is $\Delta = 2.7 \, \lambda_{D}$ where $\lambda_{D}$ is the Debye length of background electrons and the grid number is [512, 512]. The simulation time and domain are 2000 $\omega^{-1}_{pe}$ for Cases W and M, and 2500 $\omega^{-1}_{pe}$ for Case S, and the simulation domain in space is given by $Lx = Lz = 25 \, c/\omega_{pe}$. The resolved ranges of wave number and frequency are then [-16, 16] $\Omega_{ce}/c$ and [0, 3.2] $\Omega_{ce}$, respectively. In each cell for each species, we employ 1000 macroparticles. The charge neutrality condition is maintained.

Note that VDFs of energetic electrons within the loop change with many factors, such as time, location, form of initial distribution, location of injection, scattering effect, etc. This results in lots of variations. Keeping this in mind, the present study is taken as a starting point along the proposed line of research, with two main aims: (1) how VDFs evolve along the loop structure while considering various levels of scattering, and (2) whether and how VDFs are capable of efficient excitation of harmonic X modes (X2, as well as other wave modes) so as to partially address the escaping difficulty associated with ECME. In the following, we first present evolution of VDFs within four different sections of the loop, then compare outputs of PIC simulations that are initiated by three sets of VDF obtained at the same location (loop top) and time (1s after injection) with different levels of scattering. Results with VDFs at other location and time, as well as using different forms of initial distribution within a different loop structure will be examined in future.

\section{Numerical Results and Analysis}
\subsection{VDFs given by the GC simulation}
In our GC simulation, motions of two millions of electrons are resolved. To evaluate evolutions of VDFs along the loop, we select four sections, referred to as A, B, C, and D, respectively (see Figure 1b). Their lengths are taken to be large enough ($\sim$12 arcsec) so as to get sufficient number of electrons. The VDFs at certain moment are obtained by collecting the information of all electrons within the specific section at the moment of interest.

Note that electrons with small initial pitch angles may get lost from the loop. Electrons that are bounced backwards give rise to the well-known loss-cone distribution. If electrons are also mirrored from the other side of the loop, a double-sided loss-cone distribution is obtained. For the selected loop, the ratio of maximum to minimum $B$, i.e., the mirror ratio, is $\sim$ 3, this gives a loss cone angle of $\sim$ 35$^\circ$. When scattering effect is included, all electrons will get lost eventually with stronger scattering leading to faster loss. These deductions are verified by our GC simulations.

Figure 2(a-d) present temporal evolution of the obtained VDFs at different loop sections (A-D) for Case W with $\tau = 5s$. The double-sided loss cone appears as an overall morphology of VDFs within the first few seconds. Angle of the loss cone and timing of its development are consistent with the above estimate. The loss-cone angle increases from Section A to Section D, as a result of the decrease of mirror ratio. 
As seen from the accompanying movie for section A (around the top), the initial VDF is Maxwellian since this section partially overlaps with the injection region while lower sections include no electrons initially.

In addition to the loss-cone feature, there appear several strip-like features that exhibit positive gradient. The lower strip feature appears shortly after the injection, corresponding to particles with relatively large pitch angles and therefore small bouncing distance; the upper one appears with large $v_\perp$ and large-positive $v_\parallel$, then mitigating to regions with smaller $v_\perp$ and negative $v_\parallel$. 
The strips appear earlier for sections closer to the top. They are given by energetic electrons that have experienced bouncing by at least one time and arrive at the section window at the specific time. Similar multi-strip features have been demonstrated earlier by \citet{1983_White} in their analytical evaluation of VDFs for a one-dimensional magnetic field line. We have repeated their calculation and obtain similar results, confirming the validity of the simulation result presented here. The features are of major interest here since they emerge along the whole structure with significant positive gradients of VDFs, this makes them to be candidates of efficient driver of ECME.

We also conducted GC simulations for cases with moderate (Case M) and strong (Case S) levels of scattering. The solutions within Section A at the same time after the injection ($t = 1s$) are presented in Figures 2e and 2f (and the accompanying movie). They are basically similar to that for Case W. With stronger scattering effect, electrons are lost from the loop at a faster pace, the strip feature becomes less sharp, more blurred, and number of observable strips may decrease. For example, only very weak signature of strips remains in Case S (see Figure 2f).

\subsection{PIC simulations of ECME radiation}
The PIC simulations are initiated by the three sets of VDFs of energetic electrons that correspond to different levels of scattering, referred to as Cases W, M, and S, respectively. The VDFs are obtained at 1s after injection and have been shown in Figures 2(a, e, f). The parameter setup of the PIC simulation has been presented above. In the following texts, we first introduce results for Case W, and then compare them with other cases.

\subsubsection{PIC simulation for Case W}
We start from Case W which corresponds to weak level of scattering with $\tau$ = 5s. The PIC-evolved distribution has been presented in Figure 3a ($t\sim1000 \,\omega_{pe}^{-1}$) and the accompanying movie. After $t\sim500 \,\omega_{pe}^{-1}$, energetic electrons start to diffuse significantly and the strip-feature gets blurred, indicating strong wave-particle interaction.

In Figure 4a we present the $\vec k$-space intensity distribution of the strongest wave mode at the corresponding wave vector $\vec k$, the $\omega$-$k$ dispersion analysis is presented in Figure 5a and the online animation. We see that the strongest mode is the quasi-perpendicular ($90^{\circ}<\theta_B<110^{\circ}$) Z mode with frequency slightly less than $\Omega_{ce}$, while the harmonic X mode (X2) is also at a high intensity along the quasi-perpendicular direction ($95^{\circ}<\theta_B<105^{\circ}$) with frequency slightly less than $2 \,\Omega_{ce}$. The X1 emission is very weak and mainly around $30^{\circ}$. Energy curves of these modes have been plotted in Figure 3d. From the energy curves, X2 and Z modes first grow linearly at basically the same growth rates which are fitted to be 0.033 and 0.026 $\omega_{pe}$, respectively (see Figure 3d for fitting lines); before $t\sim500 \,\omega^{-1}_{pe}$, the two modes reach the maximum intensities of $\sim2.9 \times 10^{-3}$ and $\sim4.8 \times 10^{-3}$ $E_{k_{0}}$, respectively, where $E_{k_{0}}$ represents the total kinetic energy of energetic electrons.

The VDFs exhibit two types of features with positive gradients, i.e., the loss-cone and the strip-like feature. To figure out the driving agency of each mode, we have plotted resonance curves onto the corresponding VDF map as shown in Figure 2a. The curves are for the specific parameter set of ($\omega, k, \theta$) at which the strongest intensities for X2 and Z modes are achieved, while for X1 we select a representative propagation angle of 35$^\circ$. It is clear that the resonance curves of X2 and Z modes almost overlap with each other, and both modes are excited by the same upper strip-like feature through the electron cyclotron maser instability (\citealp{1979_Wu}). The lower strip-like feature passes through the background Maxwellian distribution, therefore unable to excite any wave mode. On the other hand, the resonance curve of X1 crosses the loss-cone feature, results in a weak excitation.

In summary, for Case W we get efficient excitation of Z and X2, yet without significant excitation of X1, both Z and X2 are associated with the upper strip-like feature of the VDF, while the loss-cone feature only leads to weak excitation of X1.

\subsubsection{Comparison of PIC simulations with different levels of scattering}
In this subsection, results of PIC simulations for Cases M ($\tau=1s$) and S ($\tau=0.5s$) are presented, and differences of the three cases are highlighted. The initial and PIC-evolved VDFs have been presented in Figures 2(e-f) and 3(b-c). The corresponding $\vec k$-space intensity distribution of waves within time range of $1500-2000 \,\omega_{pe}^{-1}$ for Cases S $\&$ M have been presented in Figure 4(b-c), results of $\omega$-$k$ dispersion analysis have been presented in Figure 5(b-c) and the online animation, and energy curves for various wave modes have been presented in Figure 3(d-f).

In general, it takes longer time for modes to saturate in cases with stronger scattering. As seen from the energy curves (Figure 3), in Case M the duration of the linear stages of X2 and Z are both $\sim 800 \,\omega_{pe}^{-1}$, longer than the corresponding duration of the two modes in Case W ($\sim 500 \,\omega_{pe}^{-1}$), and shorter than the duration of the linear growth of X1 in Case S which is $\sim 1800 \,\omega_{pe}^{-1}$.

In Case M, intensities of both X2 ($\sim3.0\times10^{-5} \,E_{k_{0}}$) and Z ($\sim3.3\times10^{-4}\,E_{k_{0}}$) are much smaller than those of Case W, and X1 ($\sim2.1\times10^{-5}\,E_{k_{0}}$) becomes considerably enhanced over the noise. Ranges of frequency and wave number of X2 ($\sim1.92-1.95 \,\Omega_{ce}$) and Z ($\sim0.96-0.98 \,\Omega_{ce}$) modes are quite close to those of Case W. In Case S, X1 becomes the strongest, reaching a maximum energy of $\sim3.8 \times 10^{-4}\,E_{k_{0}}$, while both Z and X1 have energies of $\sim2\times10^{-7}\,E_{k_{0}}$) much weaker than those in other cases. X1 has a symmetric arc-like emission pattern along quasi-parallel to oblique direction ($0^\circ < \theta_B < 55^\circ$), its intensity is close to the thermal noise level for larger $\theta_B$. As read from the VDFs shown in Figures 2 and 3, densities of energetic electrons along the edge of the loss cone in Case S are much higher that those in Case W. This is due to different level of scattering. With stronger scattering the loss cone feature is better-developed within shorter period and the strip-like features are less significant, while for weaker scattering, a significant part of energetic electrons is still bouncing and the VDF is characterized by strip-like features at the time of interest (1s after injection). It is apparent that the upper strip-like feature is more efficient (than the loss-cone feature) in converting electron energy into wave modes. This explains why the loss-cone feature in Case W and Case S plays very different role and why the duration of linear stage in weaker-scattering case lasts shorter.

Resonance curves shown in Figures 2e and 2f are plotted with parameters provided in the figure caption. Again, Z and X2 modes (if exist) are excited by the upper strip-feature in Cases M $\&$ W, and the very-strong emission of X1 in Case S is driven by the loss-cone feature. The linear growth rate of X1 in Case S is much smaller than that of X2 and Z in Cases W and M. Note that similar emission pattern of X1 via the loss-cone maser has been reported by earlier studies (see, e.g., \citealp{1983_Wagner},\citealp{1984_Wagner}; \citealp{1995_Yoon}).

\section{Summary and Discussion}
As a starting point to bridge the large-scale dynamics and small-scale plasma kinetic maser instabilities associated with solar radio bursts, in particular, solar spikes, we developed a numerical scheme combining techniques including the magnetic field extrapolation to describe the magnetic configuration of a normal loop, the guiding-center method to infer the temporal evolution of VDFs along various sections of the loop while taking the effect of pitch-angle scattering into account, and PIC simulations to further explore the kinetic instabilities driven by electrons with the obtained VDFs. Consistent with earlier studies, VDFs of energetic electrons that are released from the loop top manifest interesting strip-like feature with significant positive velocity gradient, together with the well-known loss-cone feature. According to further PIC simulations, the strip-like feature is essential for efficient excitation of ECME at harmonic X mode (in the weak-scattering case), while the loss-cone feature can be efficient in exciting the fundamental X mode (in the strong-scattering case). Efficient amplification of X2 favors the escape of ECME radiation from the corona, this effectively reduces the limitation of applying ECME to solar spikes. The study provides new insight into how escaping radiation, in particular, the harmonic X mode, can be generated within a coronal loop.

In most earlier studies relevant to ECME and solar radio spikes (e.g., see \citealp{1990a_Aschwanden}), loss-cone distributions have been applied. This distribution may lead to efficient excitation of X1 mode, while being unable to drive efficient excitation of X2 or higher harmonics. During its outward escape, it is difficult for the X1 mode to pass the second-harmonic absorption layer where the absorption coefficient is notoriously large. This gives rise to the escaping difficulty of fundamental emission from the corona. To resolve this issue, efficient excitation of radio bursts at higher harmonics is required. Thus, the most significant point of this study is the efficient excitation of X2 under certain conditions. Here we demonstrate that strip-like features of VDFs can form due to the initial bouncing motion of energetic electrons that are released at the loop top. These features are formed during the early stage of the VDF relaxation towards a well-developed loss-cone distribution, and they are critical to efficient excitation of harmonic X mode.

It should be pointed out that excitation of ECME radiation depends sensitively on many conditions. The effect of pitch-angle scattering on both VDFs and further excitation of wave modes has been demonstrated. In addition, initial distribution of electrons, injection location and the way of injection (e.g., impulsively or continuously, see \citealp{1983_White}), ratio of the characteristic frequencies ($\omega_{pe}/\Omega_{ce}$), all these factors shall affect the obtained form of VDFs and further excitation of various wave modes. Future study shall expand the parameter regime for a more-complete understanding of the multi-scale radiation processes.

\section*{Acknowledgments} 
The present study is supported by the National Natural Science Foundation of China (11790303 (11790300), 11750110424, and 11873036). The authors acknowledge the open-source Vector Particle In Cell (VPIC) code provided by Los Alamos National Labs (LANL), Super Cloud Computing Center (BSCC, URL: http://www.blsc.cn/) for providing HPC resources, and Dr. Alexander William Degeling and Xiangliang Kong (SDU) $\&$ Dr. Mahboub Hosseinpour (TU) for helpful discussion.

\newpage
   \begin{figure}[h]
 \centering
 \includegraphics[width=0.9\linewidth]{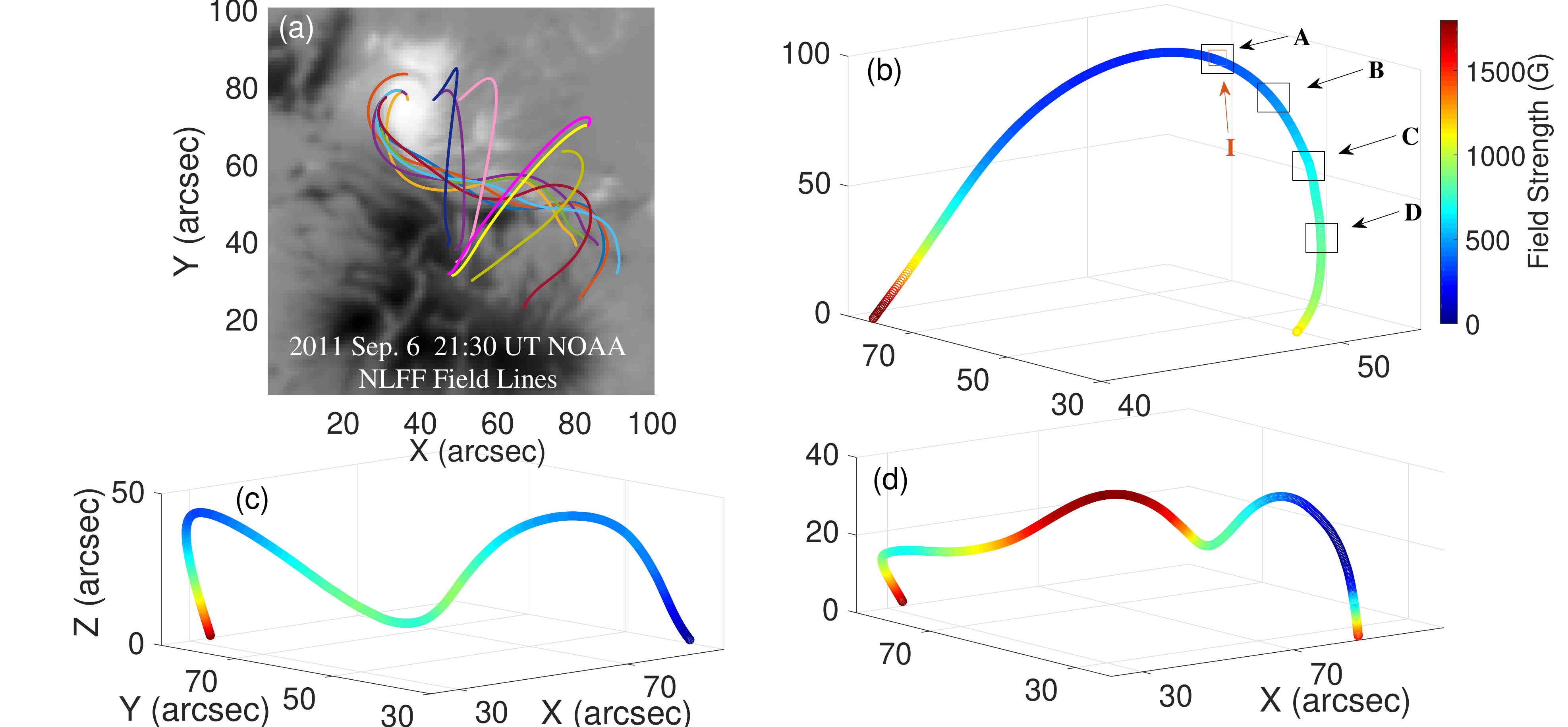}
 \caption{(a) The magnetogram of HMI for AR 11283 before the eruption of the major X2.1 flare on September 6, 2011, over-plotted with field lines given by NLFFF extrapolation; (b-d) representative lines with different amount of twist. The color in panels (b) represents the field strength of lines presented in (b-d). The letters “A-D” represent the sections within which VDFs will be examined, and “I” represents the region of injection.}
 \label{}
 \end{figure}

   \begin{figure}[h]
 \centering
 \includegraphics[width=0.9\linewidth]{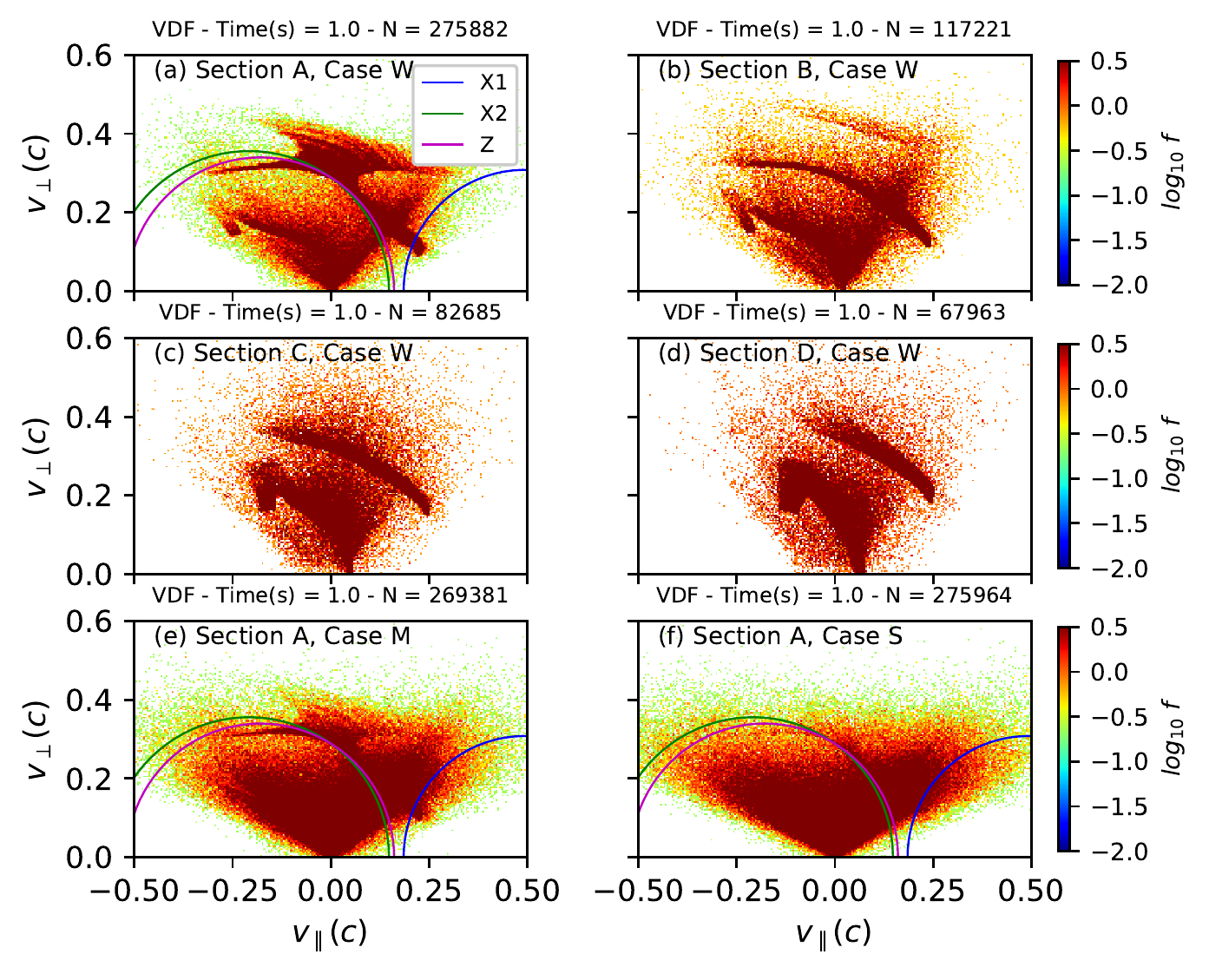}
 \caption{(a-d) The obtained VDFs at a specific time (1s after injection) within different loop sections (see Figure 1b) for the case with weak level of scattering, i.e., Case W with $\tau = 5s$. Numbers of particles ($N$) used to get the integral VDFs are written on top of each panel. (e-f) VDFs corresponding to moderate (M), and strong (S) levels of scattering for Section A, also obtained at 1s after injection. Resonance curves corresponding to the X2, X1, and Z modes are over-plotted. For X2, parameters used to plot these curves are (the wave frequency $\omega = 1.98 \,\Omega_{ce}$, the total wave number $k= 1.92 \,\Omega_{ce}/c$, the propagation angle $\theta = 100^\circ$, and the harmonic number $n=2$); for X1 they are ($1.08 \,\Omega_{ce}$, $0.65 \,\Omega_{ce}/c$, $35^\circ$, and 1), for Z they are ($0.96 \,\Omega_{ce}$, $1.15 \,\Omega_{ce}/c$, $100^\circ$, and 1).
The animation of this figure (panels (a), (e) and (f)) for the main three cases (W, M and S) begins at t = 0s and advances 0.1s at a time up to t = 3s then continues with 2s interval till the end, at t $\sim$ 19s. The realtime duration of the video is 7s.\\ 
  (An animation of this figure is available).}
 \label{}
 \end{figure}

   \begin{figure}[h]
 \centering
 \includegraphics[width=0.9\linewidth]{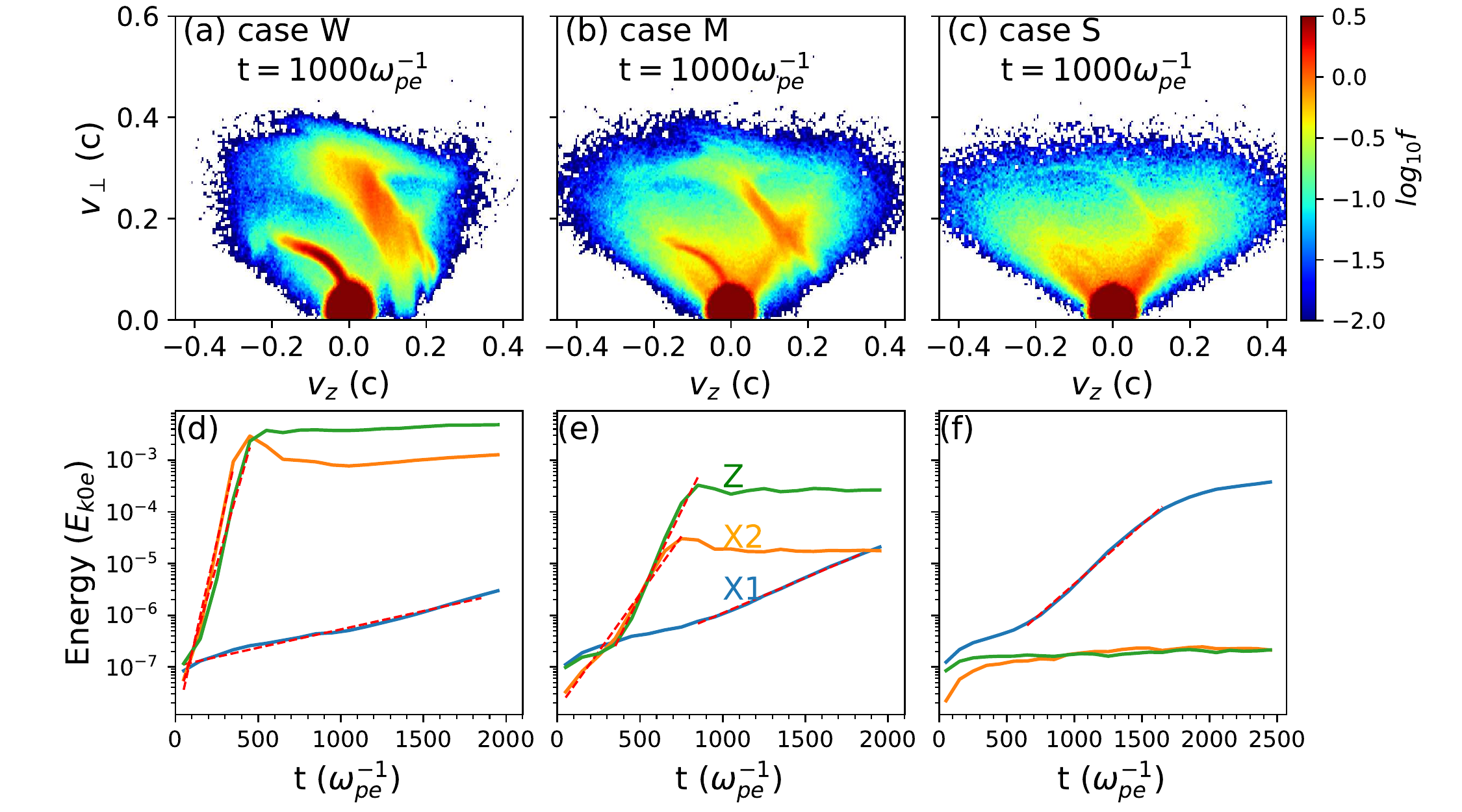}
\caption{Upper panels: VDFs at $t = 1000 \,\omega_{pe}^{-1}$ obtained by the PIC simulation for Cases W, M, and S; lower panels: Temporal profiles of energies of various wave modes (X2, X1, and Z), normalized to the total energy of energetic electrons ($E_{k_{0}}$). Dashed lines represent exponential fitting of linear growth rates, which are (0.033, 0.0016, 0.026) $\omega_{pe}^{-1}$ for X2, X1, and Z in Case W, and (0.01, 0.0031, 0.015) $\omega_{pe}^{-1}$ for X2, X1, and Z in Case M, and 0.0053 $\omega_{pe}^{-1}$ for X1 in Case S, respectively. Energy profiles are calculated within squares plotted in Figure 4b.
The video begins at $t = 0\,\omega_{pe}^{-1}$ and advances $100\,\omega_{pe}^{-1}$ at a time up to $t = 2000\,\omega_{pe}^{-1}$. The realtime duration of the video is 4s.\\ 
 (An animation of the upper panel of this figure is available).}
 \label{}
 \end{figure}

   \begin{figure}[h]
 \centering
 \includegraphics[width=0.9\linewidth]{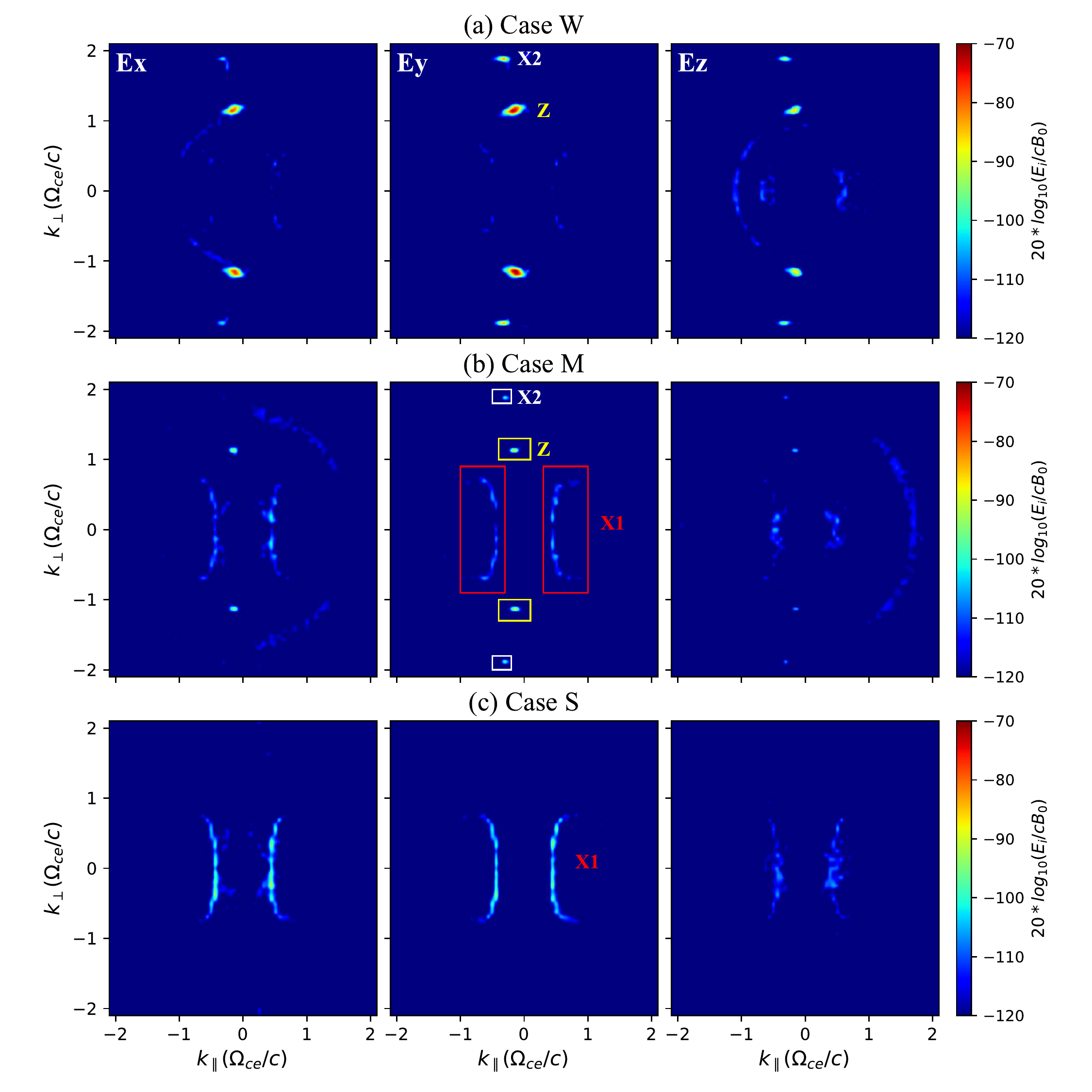}
 \caption{Distribution of maximum intensity of electric field components in the $k_{\parallel}$-$k_{\perp}$ space as shown by the color-map of 20 $log_{10}$ $[(Ex, Ey, Ez)/(cB_{0})]$ for (a) Case W, (b) Case M, and (c) Case S, with left panels for $E_x$, middle panels for $E_{y}$, and right panels for $E_{y}$. The time interval of analysis is 1500-2000 $\omega_{pe}^{-1}$, “X2”, “X1”, and “Z” stand for the harmonic X, fundamental X, and Z mode, respectively. Squares in panel b represent the areas used to calculate energy of respective wave modes. See Figure 3 for the obtained energy profiles.}
 \label{}
 \end{figure}

   \begin{figure}[h]
 \centering
 \includegraphics[width=0.9\linewidth]{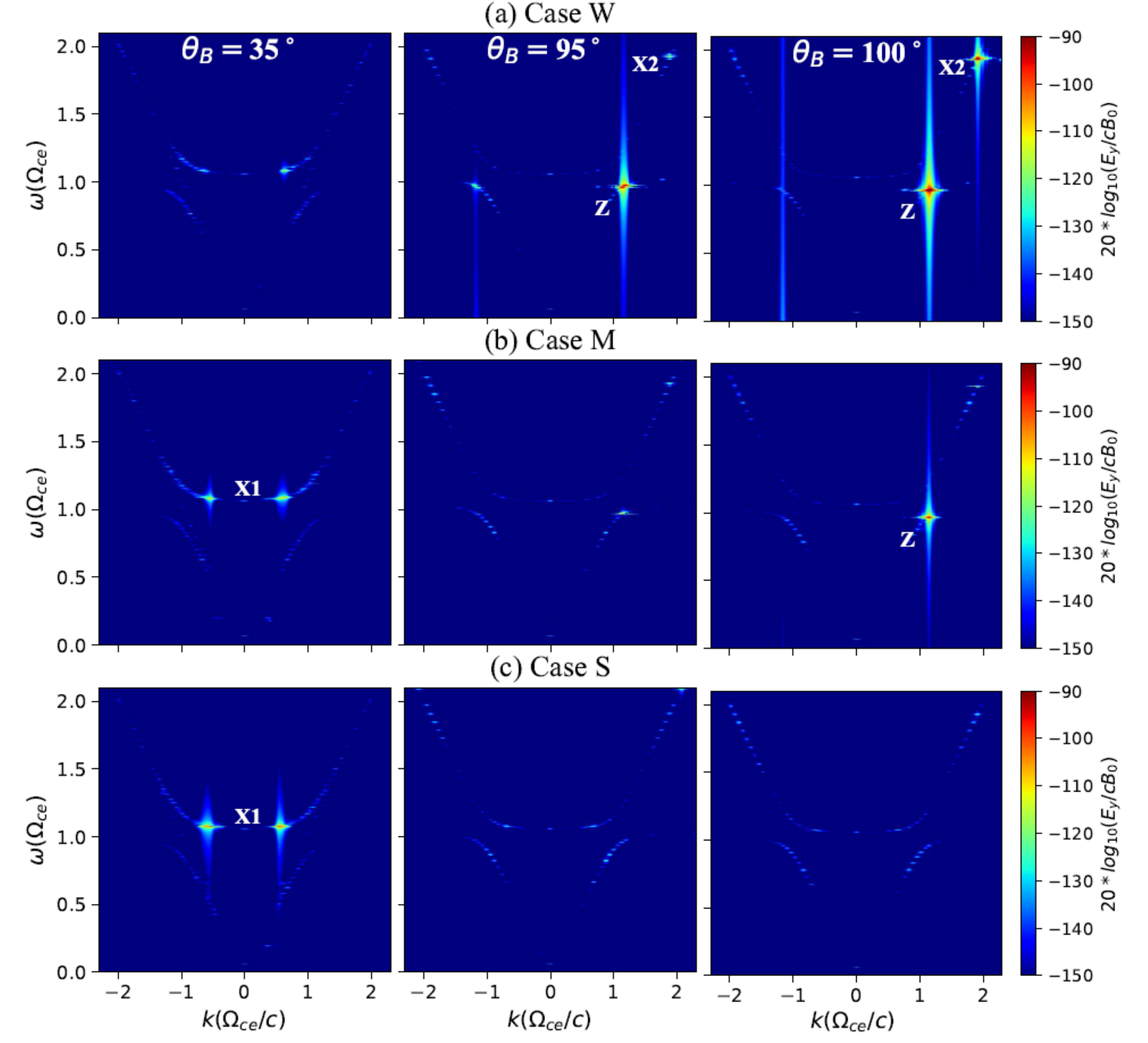}
 \caption{Wave dispersion diagrams for (a) Case W, (b) Case M, and (c) Case S at three propagation angles ($\theta_{B}= 35^{\circ}, 95^{\circ}$, and $100^{\circ}$). The time interval of analysis is taken to be 1500 $< \omega_{pe}^{-1} <$ 2000, “X2”, “X1”, and “Z” stand for the harmonic X, fundamental X, and Z mode, respectively. 
The video begins at $\theta_{kB} = 0^{\circ}$ and advances $5^{\circ}$ at a time up to $\theta_{kB} = 180^{\circ}$. The realtime duration of the video is 7s.\\
(An animation of this figure is available).}
 \label{}
 \end{figure}

\end{document}